# A NOVEL APPROACH FOR E-PAYMENT USING VIRTUAL PASSWORD SYSTEM

Vishal Vadgama[1], Bhavin Tanti[1], Chirag Modi[2], Nishant Doshi[2]

[1]M.S. Student, BITSPilani, India

[2]Reasearch Scholar, NIT Surat, India

vvadgama@hotmail.com,

{bhavintanti, cnmodi.956, doshinikki2004} @gmail.com

## ABSTRACT

*In today's world of E-Commerce everything comes online like Music,E-Books, Shopping all most everything is online. If you are using some service or buying things online then you have to pay for that. For that you have to do Net Banking or you have to use Credit card which will do online payment for you. In today's environment when everything is online, the service you are using for E-Payment must be secure and you must protect your banking information like debit card or credit card information from possible threat of hacking. There were lots way to threat like Key logger, Forgery Detection, Phishing, Shoulder surfing. Therefore, we reveal our actual information of Bank and Credit Card then there will be a chance to lose data and same credit card and hackers can use banking information for malicious purpose. In this paper we discuss available E-Payment protocols, examine its advantages and delimitation's and shows that there are steel needs to design a more secure E-Payment protocol. The suggested protocol is based on using hash function and using dynamic or virtual password, which protects your banking or credit card information from possible threat of hacking when doing online transactions.*

## 1. INTRODUCTION

E payment is a subset of an e-commerce transaction to include electronic payment for buying and selling goods or services offered through the Internet. Generally we think of electronic payments as referring to online transactions on the Internet, there are actually many forms of electronic payments. As technology developing, the range of devices and processes to transact electronically continues to increase while the percentage of cash and check transactions continues to decrease. In the US, for example, checks have declined from 85% of non-cash payments in 1979 to 59% in 2002, and electronic payments have grown to 41%.The Internet has the potential to become the most active trade intermediary within a decade. Also, Internet shopping may revolutionize retailing by allowing consumers to sit in their homes and buy an enormous variety of products and services from all over the worlds. Many businesses and consumers are still wary of conducting extensive business electronically. However, almost everyone will use the form of E Commerce in near future.

The e-payments are stored and then converted to digital type. This will cause new difficulties during the developing secure e-payment protocol. The payment is simply be duplicated against the conventional physical paying methods. As the digital payment is characterized as simple sequences of bits, nothing in them stops them copying. As we know to do E-Payment we have to provide our debit or credit card information. At this time if our debit or credit card information is stolen by hackers than this information is misused for many purpose. Therefore, the process of E-Payment must be secure and confidential. Some website use well known secure channel and protocols like SSL/TLS to protect debit or credit card information but most commercial websites still rely on poor protection mechanism. Attacks to which secure channel authentication approach is still vulnerable are listed below.

*Forgery Detection:* Where some secret key like public or private key is detected by hackers during transaction. If the formula to calculate this is very easy and information can be get easily.

*Shoulder Surfing:* Here attacker use a hidden camera to record all keyboard actions of a user and these actions on a keyboard can be studied later to figure out user's credit card information like account no and password. This attack most likely occurs in insecure and crowded public environments such as Internet cafe, shopping mall, airport, etc.

*Key logger:* This attack contains or installs some malicious code in machines, which capture keystrokes and stores them somewhere in the machine or send them back to the adversary. This attack provides the adversary with any strings of texts that a person might enter online like credit card information.

*Phishing:* This attack proven to be very effective, will acquires sensitive information, such as credit card information, by masquerading as a trustworthy person or business in an electronic communication. For example, a phisher can set up a fake website which will sell things at cheapest price ever and then send some emails to potential victims to persuade them to access the fake website.

Users who don't aware of this fake website will go to this website and gives credit card information to buy things. This way, the phisher can easily get all information of credit card. From all of this fact we have noted that a password should be dynamic to prevent above attacks. It is very reasonable that a password should be constant for the purpose of easily remembering it. However the price of easy to remember is that the password can be stolen by others and then used in e-payment. Also we should able to set maximum amount which is transferred during one transaction, so that if our debit or credit card information is stolen by attackers, we will not loss all of our money and instead loss only money which is set here.

**1.1 E-payment protocol**

The E-Payment protocol is a protocol, which solves the problems related to old protocols by introducing a new concept. Our protocol strongly follows characteristics like Anonymity, Divisibility, Transference, Over spending detection. Through using our protocol you can get anonymous cash Transaction and efficiency at same time. Our protocols is protected by dynamic or virtual password that involves a small amount of human computing in an internet based environment, which will be resistant to attacks like Phishing scan, a Trojan horse, shoulder-surfing, etc. We purpose a virtual password concept that requires a small amount of human computing to secure user's password concept. We will adopt user-determined randomized linear generation function to secure user's temporary password.  We are going to adopt user-determined randomized linear generation functions to secure users passwords based on the fact that a serer has more information than any adversary does. For this protocol when company register for a new account, Bank or Credit card provides him a secret code which will generate temporary password and random number for user. With the real password and credential information user can set the limit balance for the withdrawal from account. This will be done a web service provided by bank of credit card providers. Now once user will get information about random number, password and user name,an user can use this temporary credentials to buy an company account in our information. Therefore, now that account only allow up to a limited balance withdrawal from the user account. User already sets the limit so if anything wrong happens that at max user can losses only up to limit balance from his account. If user requires more money online then they can generate more random number and password as given by application. Therefore, at any point in real time transaction user will not going to reveal his original information so his/her account information will not be compromised.

*Organization of the paper:* The rest of the paper is organized as follows. We describe related work about done so far in chapter 2. In chapter 3, we propose the virtual password scheme. In chapter 4, we describe the proposed work. Finally, we conclude our paper and describe our future work.

## 2. LITERATURE SURVEY

### 2.1. Virtual Password

The authors in [7] introduce the concept of the E-cash since than it is still hot favorite topic for research. In [8] [9] authors give the each system in single condition support. In [2] authors give design of E-cash system based on offline concept. In [3] authors discuss model based on java technology. In [1] authors used the concept of virtual password for email login system.

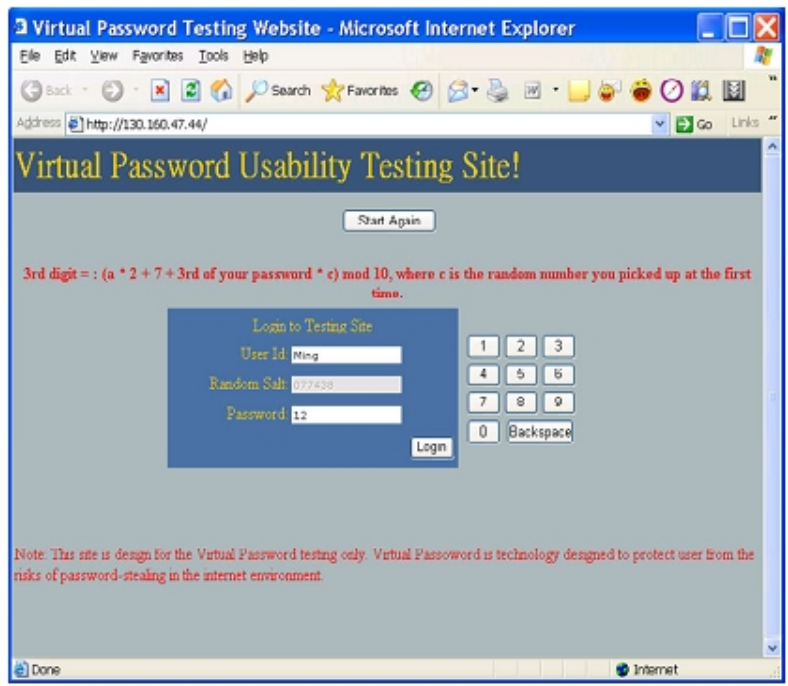

Figure 1 : Login Page[1]

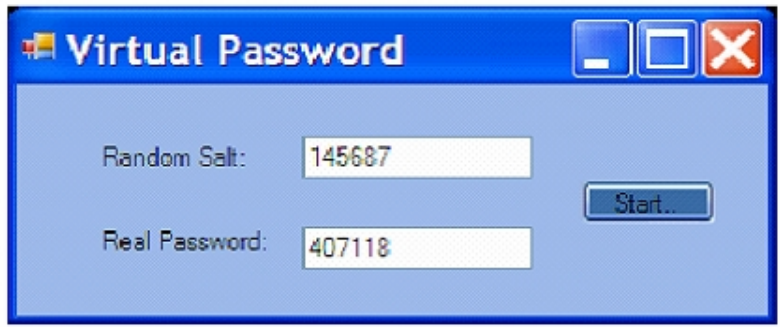

Figure 2: helper application for PC [1]

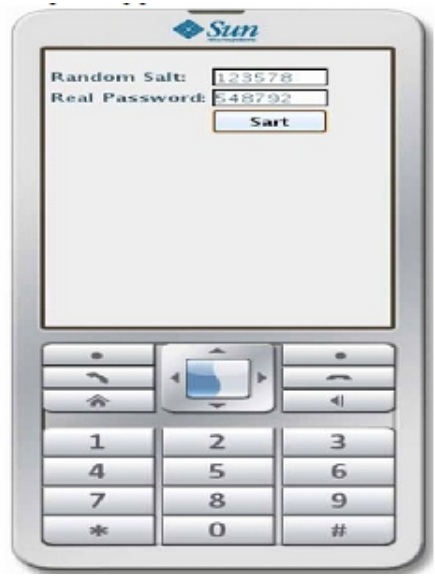

Figure 3 : Helper application for Mobile [1]

In this paper we had extend this concept to E-cash system. In [4] authors used the concept of blind signature concept. In [5] [6] authors extend the E-cash for faster and efficient in terms of efficiency. In [10] authors used the concept of zero knowledge for E-cash system. In [11] authors proposed E-cash for offline support. In [12-17] authors proposed the E-cash with security and several variations. But as we observed in all of the previous approach, when any communication or secret of customer is compromised than attacker can transfer all money from victim's account. In our approach even the temporary password of customer compromised, then attacker only transfer money up to certain limit, so giving privacy to customer's account more privacy and anonymity.

During E-Payment process bank gives us all details to do E-Payment, like user-id and password. We will enter this password during E-Payment process. Here we will enter same password each time we do transaction, so we can call this password a static password. But it is dangerous to use static password during E-Payment, because hackers steal once this password he can use this password for E-Payment or other purpose. Therefore, we need to use a password, which will change each time when we do transaction, but it's not possible to get new password each time from bank. Therefore, we will use a password called dynamic or virtual password, which will change during each transaction. A dynamic or virtual password is a password gendered from static password and some random number using virtual function. Therefore, user first generates a dynamic password and this generated password is submitted during actual e-payment process. Therefore, if hackers steal this password he can only able to use this in one transaction in next transaction he need to give new password, which is not easy to generate without virtual function.

### 2.1.1. Virtual Password Approach

A virtual password P is composed of two parts, a fixed alphanumeric A and a function F. Password P can be computed by P=(A, F) and function F can be computed by F (A, R) = Pd, where R is a random number and Pd is a dynamic password used for authentication. Since we call P=(A, F) a virtual password, we call F a virtual function. The user input includes (ID, Pd), where ID is user ID and Pd is virtual password. On the server side, the server can also calculate Pd in the same way to compare it with the submitted password.

In the virtual password approach, the registration process is similar to other traditional approaches i.e. user enters user ID and password; the only difference is that user needs to send virtual function during registration phase. The server at the end of registration process delivers this function information to the user. The user needs to remember this function together with the password they have chosen. The user-specified password and the system-generated function are combined into a virtual password. We can say that virtual password approach is resistant to directory attack, where users creates password either related to their own name, birth date, other simple words, etc. In virtual password approach user can change virtual password also, different from the traditional scheme, users can change the fixed part of the virtual password or the virtual function, or even both.

### 2.1.2. Virtual functions

To generate virtual password user needs to use virtual function. Virtual function takes two inputs, one is fixed password and second is random number. Here fixed password is a password, which user has entered during registration process and random number is a number prompted on login screen by server. Virtual function should be unique per user, so that a user cannot know the computation process of virtual password of other user. This function is provide by bank to user through email or prompted on screen. Virtual function should be simple so that user can remember it. But for better password protection it should be hard. Bank can provide a helper application to user, which use this virtual function and generate password for user. If this application is not available to user, user needs to calculate it manually.

The virtual function plays a critical role in the virtual password. There are an infinite number of virtual functions, so that designing an appropriate function is very critical to the success of our scheme. If a helper-application application is available for the user, the user needs to type the random salt into the helper-application, and subsequently, the dynamic password is generated by the helper-application. Then the user types in the generated dynamic password in the login screen. In this way, the extra time required is very small and the precision will be one hundred percent correct as long as the user types the correct random salt displayed on the login screen.

If there is no helper-application for a user, the user needs to calculate the dynamic password from the virtual function with the inputs, random salt and the fixed part of the virtual password. The whole login process may take a little bit longer because it requires the user to perform some calculations. This must work for the user who has no mobile device, so in that case, the virtual function should not be too complicated for human computing.

## 2.2. E-cash Payment

Many protocols are designed to do secure E-Payment and are listed below. In [18] authors proposed system which is relied on a single use token method. The user creates blinded e-bank currency note and passes it to the bank to be signed using bank public key. The bank signs the currency note, subtracts the value from the user account, and returns the signed currency note back to the user. The user removes the blind thing and utilizes it to buy goods from the super market. The super market checks the authenticity of the bank currency note using the bank public key and passes it to the bank where they are verified contrary to a list of currency note already used. The amount is deposited into the supermarket account, the deposit approved, and the supermarket in turn emits the merchandise.

In [19] authors proposed system which is a decentralized e-payment protocol, and it allow payments as low as 1/10 of a cent. It employs a type of e-coins. It is introduced to make the cost of committing a fraud, more than the cost of the real transaction. It utilizes asymmetric encryption techniques for all information transactions. Millicent is a lightweight and secure scheme for e-commerce through the Internet. It is developed to support to buy goods charging less than a cent. It is relied on decentralized validation of e-currency at the seller server without any further communication, costly encryption, or off-line processing. In [20] authors suggested method in which there is a possibility to reduce the number of messages engaged with every transaction. Also, the lottery ticket scheme is relied on the assumption that financial agents are risk neutral and will be satisfied with fair wagers. In [21] authors suggested methodwhere e-vouchers can be moveable but the direct exchange between purchasers and vendors is impossible. As a result, a financial agent is needed and this will raise the transactions charges of exchange.

In [22] authors suggested method inwhich supports multiple merchants is an efficient and flexible protocol. The protocol is divided into three schemes: certificate issuing scheme, payment scheme, and redemption scheme. However it contains some limitations in his scheme like the system performance is reduced by necessarily frequent signing in each transaction, the customer has to keep different hash chains and corresponding in-dices, however the overhead of merchants is relatively high. To securely deposit, the bank has to collect all pay-words belonging to the same chain. It needs an additional storage space and wastes undetermined waiting time and the dispute arises if the merchant forges transaction records or the customer double spends. However scheme in [22] is most popular than others some limitations is observed by [23] like the system performance is reduced by necessarily frequent signing in each transaction, the customer has to keep different hash chains and corresponding indices; however the overhead of merchants is relatively high. To securely deposit, the bank has to collect all pay words belonging to the same chain. It needs an additional storage space and wastes undetermined waiting time, and the dispute arises if the merchant forges transaction records or the customer double spends. The protocols with idea of public key encryption scheme using the thought of hash chain is presented by [23] and has blind scheme using RSA-typed blind signature which overcomes limitations of Kim and Lee protocol.

As we have seen many schemes, protocols and software have been designed to prevent e-payment from possible threat of hacking. However to the best of our knowledge, so far, there is not a scheme, which can defend against Shoulder-surfing, Key logger and Phishing at the same time. In E payment protocol we will try to defend all of above attacks.

## 3. PROPOSED PROTOCOL FOR E-CASH

We have seen that when using above protocol, user has to send his secret information like debit or credit card information to bank to get the certificate. We know banks use some secure protocol like SSL/TLS for transmitting private data over the web is well-known in academic research, but most current commercial websites still rely on the relatively weak protection mechanism of user authentications via plaintext password and user ID.

[1]has provided good concept of virtual password in his paper, which we can use here to secure our E-Payment from possible threat of hacking. [1]suggested using virtual password, which we have seen in above chapter. Virtual function is required to generate virtual password. [1]have first suggested using following Linear Function.

$$B(x_i) = [a(x_i + y_i) + c] \bmod Z \quad (1)$$

where a and Z are relatively prime, $x_i$ is one digit from the fixed part of the user's virtual password, $y_i$ is one random digit provided by the system, and a and c are the constant factors of the linear function, which the user has to remember. The $B(x_i)$ is a bijective function if and only if a and Z are relatively prime.

[1]showed that using above function we can protect user password from Phishing, key logger, and shoulder-surfing attacks but not from multiple attacks. [1] showed that if user is lured to try to login to any same phishing website more than twice, it will leak his/her password. Reason for that is for any given ith digit of the fixed part of users virtual password, if user has tried more than twice to login to the fake website, then adversary could obtain the two equations below: $[a(x_i + y_i) + c] \bmod Z = k_i$, and $[a(x_i + y'_i) + c] \bmod Z = k'_i$. Now Bob can know that $[a(y'_i - i)] \bmod Z = (k'_i - k_i) \bmod Z$, and as a result, it can calculate a. After a is identified by the adversary, the system is broken. Then adversary can use user's account to login to the real website in the following way. For the ith digit, adversary can just type in $(k_i + a(y''_i - y_i)) \bmod Z$, where $k_i$ is the first time user typed in the ith digit in the fake website, $y_i$ is the ith random digit provided by the fake website, and $y''_i$ is the ith digit the system will display on the screen, which adversary needs to login to users account.

To overcome this limitation [1] proposed randomized linear generation function that uses the value of a digit in the dynamic password to calculate a subsequent digit in the dynamic password.

$$B(xi) = \frac{k_i = (ax_1 + y_1 + x_2 + c) \bmod z, \ i=1}{k_i = (ax_{i-1} + y_i + x_i + c) \bmod z, i=2,3 \ldots n} \quad (2)$$

Where a is a constant which the user needs to remember but c is not. The most interesting part of the function is that c will be a random number which the user randomly picks each time when the user tries to login to the system. Since gcd(a, Z)=1, the above function is also a bijective function regardless of the c value. Because c is also unknown to server, the server knows that c ∈{0,1…Z-1}. The authentication could be done as follows.

Let B$^{-1}$(x) be the reverse function of B(x). After the server gets the user's keyed dynamic password k1,k2...kn, and the fixed part of the virtual password of the user, x1,x2…xn, the server can perform the following verification:

Verify ()
{For each digit u∈{0,1...Z-1} {
For each digit in the dynamic password the user typed
{wi =B$^{-1}$(ki,u)}
if (w1w2…wn= x1x2…xn) return true}
Return false
}

The algorithm above guarantees that if the user has input the correct password, the system will grant him/her entrance whatever the random number he/she picked. However, it is also true that for each user, there will be multiple (exactly Z) acceptable dynamic passwords existing for each specific login session. This may increase the probability that the adversary's random input happens to be the correct password. However, if the length of your password is long enough, the probability is very small, i.e., $Z/2^n$, where n is the length of the password.

A scheme with equation (2) can defend against Phishing, keylogger, shoulder-surfing, and multiple attacks as follows.This function now also secures against multiple attacks. If an adversary has tricked a user into logging into his fake website twice, the adversary obtains $k_1=(ax_1+y_1+x_2+c)$ mod Z and $k_1'=(ax_1+y_1'+x_2+c')$ mod Z, where a, m1, c, and c' are unknown to the adversary, and then what information the adversary can figure out is the (c'- c) mod $Z=(y1'-y_1)+(k_1'-k_1)$. Since the c and c' were randomly chosen by the user, (c'-c) does not provide any information. If the adversary cannot work out some clue about the first digit of the dynamic password $k_1$, he/she cannot find about $k_2$ and the later digits in the dynamic password. Therefore, using the linear function with a random number can remove the possibility of multiple dynamic password attacks. We slightly modify this concept to get more secure E-Payment.

With the real password and Credential information user can set the limit balance for the withdrawal from account. This will be done a web service provided by bank of credit card providers. Now once user will get information about random number, password and user name an user can use this temporary credentials to buy an company account in our information. Therefore, now that account only allow up to a limited balance withdrawal from the user account. User already sets the limit so if any thing wrong happens that at max user can losses only up to limit balance from his account. If user requires more money online then they can generate more random number and password as given by application.

To get more security on E-Payment protocol we will use secure HTTP channels, which will transfer data in encrypted mode. In sattars e-payment protocol many information transferred from one end to another end many times. This is the point where hackers can play role, they can able to intercept this information and by every time inspecting this values they can get the logic of formula used by bank or user and can easily hack the system. Therefore, here instead of passing the information from one end to other end we can save this information's and return this information whenever required.

We first modify the actual information by using predefined formula and then save or pass this information using RC5-32 encryption system. We can make a application which will do this work and can installed on both side. On opposite this application first decrypt this information and then apply predefined formula to calculate actual information.

Also we will not save any information on client side, we will save all the information on secure database created on bank side. Therefore, hackers can have less change to hack this information ever. The information calculated from client side is passed using our application to bank side so bank will save this information. And bank does not need to call on client side to get this information they have all the required information on his side. Here this application is unique per user so that they can't get information processing logic of other use. Therefore, the suggested modified protocol is shown as below.

Step 1: Bank

1.1. Select secretly and randomly two large prime p and q
1.2. Calculate modulus $n_B = p * q$
1.3. Compute $\theta(n) = (p-1)(q-1)$
1.4. Choose exponent key e where $1 < e < \theta(n)$ and $gcd(e, (\theta(n)) = 1$

1.5. Calculate private key w where e *w = 1mod $\theta$(n)
1.6. Determine the public key (e, $n_B$) and private key (w, $\theta$ (n), p, q)

Step 2: User
2.1. Get required information from bank using suggested application.
2.2. Select arbitrary numbers r and u
2.3. Calculate a = $r^e$ * h($x_0$)( $u^2$ + 1) mod $\theta$(n)
2.4. Save (b, a) using suggested application on bank side.
Note that information b can indicate the expiry date; the value of cash (higher limit) that the user can employ that is the funds of every hash currency.

Step 3: Bank

3.1. Select an arbitrary number $x_1$<$\theta$(n)
3.2. Save $x_1$ using suggested application.

Step 4: User
4.1. Get required information from bank using suggested application.
4.3. Choose an arbitrary value r1
4.4. Calculate $b_2$= r * r1
4.4. Calculate $\beta$ = $(b_2)^e$ * (u - $x_1$) mod $\theta$(n) and save it using using suggested application.

Step 5: Bank
5.1. Calculate $\beta^{-1}$ mod $\theta$(n)
5.2. Compute $t_1$ = h $(b)^w$ * (a $(x_1^2$+1) * $\beta^{-2}$) 2*w mod $\theta$(n)
5.3. save ($\beta^{-1}$, $t_1$) using suggested application.

Step 6: User
6.1. Get required information from bank using suggested application.
6.2. Calculate $c_1$ = (u* $x_1$+1) $\beta^{-1}$ * $(b_2)^e$=(u* $x_1$+1) (u - $x_1$)$^{-1}$ mod $\theta$(n)
6.3. Calculate $s_1$ = $t_1$* $r^2$ * $(r_1)^4$ mod $\theta$(n)

Here also this application is lightweight they can give much load or performance problem. Using this we can hide our actual information from hackers and do secure transactions. [23]has used same transaction scheme of pay-word. Where root value of pay-words is merged with si that obtain from the bank B, which enables the user to employ the rest of the unspent pay-words in chain for multiple payments to other merchants. But her if this Si is stolen by hackers for any transaction than he can hack other transactions also.Therefore, here instead of using this old $S_i$ is value we have to use new value given by bank each time we do transaction for different merchant. Also we use hash key and function to get this new value. This will secure our E-payment protocol so that we cannot loss all of our money and only loss a part which is hacked by hacker in particular transaction.

Also in [23] protocol it is too time consuming for multiple merchants, so here we can use hash function and also store this function on server side and use same logic of virtual function to protect this.For this protocol when person registered for new account, bank gives him a secret code and general application which can generate temporary password and random number for user. With the real password user can set the limit balance for the withdrawal from account. Now user enters the secret number in the application and gets a random number and associated password for that random number. Now user will enter his user name, random number and the password as given by the application. Therefore, now online account only allow the up to limit balance to be withdrawal from the user account. Now if by chance random number and password for it is compromised than at max user can loss only up to limit balance from his account. If user require more money than the limit balance than he can generate more random number and password from the application. Therefore, at any point during the transaction user do not enter his real password so his account never compromise. The application can be general so it can install even on the supported mobile also or in personal PC of a user.

## 4. CONCLUSIONS

In this paper, we discussed how to prevent E-Payment system from possible threat of hacking. We have shown that our protocol is secure in multiple transaction and also do not use large key's so it is also less time consuming. We proposed a virtual password concept involving a small amount of human computing to secure users' passwords in on-line environments. We adopted user-determined randomized linear generation functions to secure users' passwords based on the fact that a server has more information than any adversary does. We analyzed how the proposed scheme defends against Forgery Detection, Phishing, key logger, and shoulder-surfing attacks.

## References

[1] M. Lei, Y. Xiao, S. V. Vrbsky, C.-C. Li, and L. Liu, "A virtual passwordscheme to protect passwords," in Proceedings of IEEE InternationalConference on Communications (ICC'2008). IEEE, 2008, pp. 1536–1540.

[2] Liu Feng; Li Xueyong; GaoGuohong; The design of an E-cash system. ICCDA 2010. Volume 2:119-222.

[3] Sarker, K.M.; Jahan, I.; Rahman, M.Z.; Secure E-cash model using Java based smartcard. ICCIT '09 ,IEEE : 626-631

[4] Sébastien Canard and AlineGouget, Multiple Denominations in E-cash with Compact Transaction Data. LNCS, 2010: 82-97.

[5] Canard, Sébastien and Delerablée, et. Fair E-cash: Be Compact, Spend Faster. LNCS: 5735, page 294-309.

[6] Canard, Sébastien and Gouget, Aline and Traoré, Jacques. Improvement of Efficiency in (Unconditional) Anonymous Transferable E-cash. LNCS 5143, 2008:202-214.

[7] D. Chuam, "Blind signatures for untraceable payments," in Advances in Cryptology - Proceedings of CRYPTO '82. New York: Plemum, 1983, pp. 199-203.

[8] L. Shi, B. Carbunar and R. Sion, "Conditional E-cash," in Financial Cryptography and Data Security '07, Lecture Notes in Computer Science, vol. 4886. Berlin: Springer, 2008, pp. 15-28.

[9] M. Blanton, "Improved conditional e-payments," in Applied Cryptography and Network Security 2008, ser. Lecture Notes in Computer Science, vol. 5037. Berlin: Springer, 2008, pp. 188-206.

[10] Okamoto T, Ohta K. Disposable Zero-Knowledge Authentication and Their Application to Untraceable Electronic Cash. In: Proc of Advances in Crypotlogy-CRYPT0l989. New York: Springer-Verlag 1990, 435:481-496.

[11] Brands S, An Efficient Off-Line Electronic Cash System Based on the Representation Problem. www.cwi.ni/fip/CWIreports/AAICS- R9323.pdf 2009,01-16.

[12] Pengtao Liu, Chengyu Hu. A constant-size linkable ring signature scheme for E-cash protocol. ICCET 2010.Volume 1 IEEE 2010:295-298.

[13] A. Lysyanslaya, Z. Ramazan, and Y. Tauman, "Group blind digital signature: a scalable solution to electronic cash", Proc. of Financial Crypto-graphy'98, LNCS 1465, Springer-Verlag, 1998, pp.184-197.

[14] M. Li, Y. Yan, C. Ma et ai, "A revocable anonymous E-cash scheme from group signatures", 10urnal of Beijing University of Posts and Telecommunications, 2005, 28(2), 30-33.

[15] Zuowen Tan. An E-cash Scheme Based on Proxy Blind Signature from Bilinear Pairings. Journal of Computers, Vol 5, No 11 (2010), 1638-1645, Nov 2010

[16] AggelosKiayias, YiannisTsiounis, and Moti Yung. "Traceable signatures," Advances in Cryptology EUROCRYPT '04, Lecture Notes in Computer Science, 3027, Springer, 2004, pp. 571–589.

[17] Yoshikazu Hanatani1, Yuichi Komano, Kazuo Ohta, and Noboru Kunihiro, "Provably Secure Electronic Cash Based on Blind Multisignature Schemes," FC 2006, Lecture Notes in Computer Science, 4107, Springer-Verlag Berlin Heidelberg, 2006, pp.236–250.

[18] D. Chaum, A. Fiat, and M. Naor. Untraceable Electronic Cash. In Crypto '88, LNCS 403, pages319–327. Springer-Verlag, Berlin, 1989.

[19] S. Glassman, M. Manasse,M. Abadi, P. Gauthier, and P. Sobalvarro. The millicent protocol for inexpensive electronic commerce. In In Proc. of WWW4 1995.

[20] R. Rivest. Electronic lottery tickets as micropayments. In R. Hirschfeld, editor,Financial Cryptography '97, pages 307–314. Springer-Verlag, 1997. LNCS no. 1318.

[21] E. Foo and C. Boyd. A Payment Scheme Using Vouchers. In Proceedings of the Second InternationalConference on Financial Cryptography, number 1465 in Lecture Notes in Computer Science, pages103–121. Springer-Verlag, 1998.

[22] S Kim and W Lee, "A Pay-word-based micro-payment protocol supporting multiplepayments", In Proceeding of the International Conference on Computer Communicationsand Networks, pp. 609-612, 2003.

[23] Sattar J Aboud" Secure E-payment Protocol". International Journal of Security, Volume 3,Issue 5: 85-92, 2009.